\documentclass[conference]{IEEEtran}

\usepackage{graphicx}
\usepackage{flushend}

\usepackage{iftex}
\ifPDFTeX
  \usepackage[UKenglish]{babel}

  \usepackage[utf8]{inputenc}
  \usepackage[T1]{fontenc}
  \usepackage{lmodern}
\else	
  \usepackage{polyglossia}
  \setdefaultlanguage[variant=uk]{english}

  \usepackage{fontspec}
  \defaultfontfeatures{Ligatures=TeX}
\fi

\graphicspath{{img/}}

\begin{document}

\title{A Distributed System for Storing and Processing Data from
  Earth-observing Satellites: System Design and Performance Evaluation
  of the Visualisation Tool}

\author{
  \IEEEauthorblockN{
    Marek Szuba\IEEEauthorrefmark{1}\IEEEauthorrefmark{3},
    Parinaz Ameri\IEEEauthorrefmark{1}\IEEEauthorrefmark{4}
    Udo Grabowski\IEEEauthorrefmark{2}\IEEEauthorrefmark{5},
    Jörg Meyer\IEEEauthorrefmark{1}\IEEEauthorrefmark{6} and
    Achim Streit\IEEEauthorrefmark{1} \IEEEauthorrefmark{7}}
  \IEEEauthorblockA{
    \IEEEauthorrefmark{1}Steinbuch Centre for Computing (SCC)
  }
  \IEEEauthorblockA{
    \IEEEauthorrefmark{2}Institute of Meteorology and Climate Research (IMK)
  }
  \IEEEauthorblockA{
    Karlsruhe Institute of Technology\\
    Hermann-von-Helmholtz-Platz~1\\
    76344 Eggenstein-Leopoldshafen, Germany
  }
  \IEEEauthorblockA{
    E-mail:
    \IEEEauthorrefmark{3}marek.szuba@kit.edu
    \IEEEauthorrefmark{4}parinaz.ameri@kit.edu
    \IEEEauthorrefmark{5}udo.grabowski@kit.edu\\
    \IEEEauthorrefmark{6}joerg.meyer2@kit.edu
    \IEEEauthorrefmark{7}achim.streit@kit.edu
  }
}

\maketitle

\begin{abstract}
  We present a distributed system for storage, processing,
  three-dimensional visualisation and basic analysis of data from
  Earth-observing satellites. The database and the server have been
  designed for high performance and scalability, whereas the client is
  highly portable thanks to having been designed as a HTML5- and
  WebGL-based Web application. The system is based on the so-called
  MEAN stack, a modern replacement for LAMP which has steadily been
  gaining traction among high-performance Web applications. We
  demonstrate the performance of the system from the perspective of an
  user operating the client.
\end{abstract}


\section{Introduction}
\label{sec:Introduction}

The processing of satellite observations
of Earth is highly data-intensive. Many satellites produce
high-resolution images of the Earth. Many observational missions were or
have been in operation for decades. In some cases the
raw input itself may not be so large but processed,
analysis-ready data is. Either way, Earth-observing satellites can now
be considered a fully fledged source of Big Data.

One such source has been the Michelson Interferometer for Passive
Atmospheric Sounding on the ESA Envisat satellite, a
Fourier-transform spectrometer. MIPAS operated between 2002 and 2012,
and measured geotemporal distribution
in the atmosphere of more than 30 trace gasses relevant to atmospheric
chemistry and climate-change research. Data from MIPAS is stored in
several different data archives, including the Large-Scale Data
Facility (LSDF)~\cite{Garcia:2011aa} at the Karlsruhe Institute of
Technology (KIT). As of August 2015, the complete MIPAS archive at LSDF
requires around 30~TB for the storage of calibrated measurement data
released by the ESA (``level-1B data'') and around 64~TB for processed
(``level-2'') data produced from the former at KIT. Both parts will
continue to grow in the near future.

MIPAS data at LSDF consists primarily of compressed text and
PostScript files as well as some classic-format NetCDF files. Analysis
of this data involves
several difficulties: there are many different sources of input,
data is separate from at least parts of its metadata,
repeated parsing of text can be time consuming, compressed files must
be wholly decompressed on the fly, data inside classic NetCDF
files is not indexed. In short, working with file-system
MIPAS data can be quite slow.

The situation becomes even more complicated when a comparative
analysis of data from different experiments is desired, for example
from MIPAS and the Microwave Limb Sounder (MLS) on the NASA Aura
satellite. With different experiments structuring
their data in different ways, the conventional analysis approach
requires developing and running several different pipelines even
though the analysis algorithm itself remains the same. Moreover, even
when multiple experiments cover the same locations and time span,
exact coordinates and time stamps of their respective data points are
only similar, not the same --- requiring another round of data
processing in order to match them.

In light of the above we have proposed and implemented an alternative,
distributed and scalable solution, which takes advantage of Big Data
tools and methods in order to improve performance of working with
MIPAS and similar data. In the following sections we shall describe
our system and present an analysis of chosen aspects of its
performance.

The structure of the following parts of the paper is as follows. In
Section~\ref{sec:systemArchitecture} we describe the architecture of
our system, discussing each of its tiers, whereas
Section~\ref{sec:Evaluation} presents the evaluation of the system's
performance from the point of view of the user of the client
application. Section~\ref{sec:relatedWork} mentions several related
tools and systems. Finally, in Section~\ref{sec:Conclusions} we
present our conclusions and the the outlook for the project.

\section{System Architecture}
\label{sec:systemArchitecture}

The goals we set while designing our system were as follows:
\begin{itemize}
\item it should scale well as the amount of data stored in it grows;
\item it should facilitate the use of data from multiple sources;
\item the basic user interface should be easy to access and use.
\end{itemize}

\begin{figure*}
  \centering
  \includegraphics[width=0.8\linewidth]{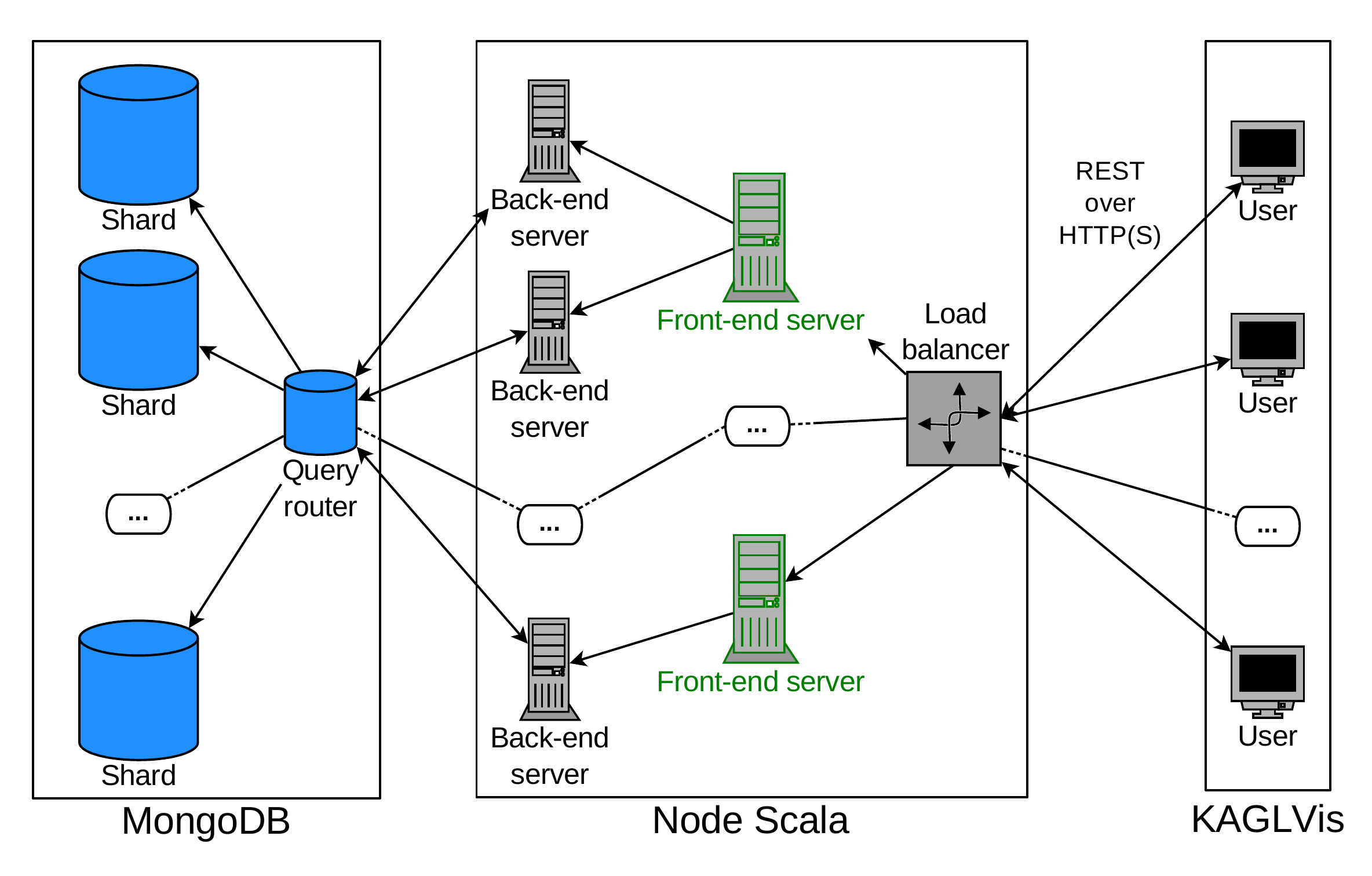}
  \caption{Architecture of our system. For clarity, the diagram omits
    internal management components of Node Scala (the controllers and
    the scheduler) as well as the Web server used to serve KAGLVis
    files to users.}
  \label{fig:systemArch}
\end{figure*}

We have chosen the standard multi-tier design model consisting of the
database, the server and the client, common among Big Data
applications. It allows for independent growth of each of the
tiers as needed as well as performing computation-intensive processing
on more powerful systems than what the client might have at their
disposal or data-size reduction closer to the database. Our system is
based on the NoSQL database MongoDB, the server-side runtime platform
Node.js, the Web-application framework Express and the JavaScript MVC
framework AngularJS, known together as the \textbf{MEAN stack} and
offering a number of advantages over more established stacks such as
LAMP to both users and developers~\cite{web:MEANstack}. The eventual
architecture of our system is shown in Figure~\ref{fig:systemArch}.

We implement the data browser as a Web application to ensure
portability and make it easier for users to keep it up to
date. Furthermore, by making our client a single-page application we
essentially add another degree of scalability to the system.

\subsection{The Database}
\label{sec:sysArchDB}

Our MongoDB system presently runs
on a single, dedicated server and contains data from
MIPAS, MLS and a number of other, smaller experiments, each with its
own database. Data for each geotemporal location constitutes a single
document.

We have developed several tools which communicate directly with the
database server using appropriate MongoDB drivers, for example
a Python tool which can be used to perform
general matching of data from two different sources (\textit{e.g.}
MIPAS and MLS); it is capable of multi-threaded operation and has
already been used to demonstrate superior performance of MongoDB
comparing to a SQL database system~\cite{Ameri:2014aa}.

\subsection{Server-side Processing and REST API}
\label{sec:sysArchServer}

Since our main client is a Web application, making our instance of
MongoDB accessible from the Internet is impractical because it would
require adding a MongoDB driver to the user's browser.
Using an intermediate server means a server-installed driver can be used,
whereas communication with clients occurs using HTTP.
Such a server can also perform pre-processing before passing the data
between the client and the database. The underlying platform of our
server is Node.js and the HTTP Representational State Transfer (REST)
client API is based on Express. Data is transferred in JSON format.

Our server is based on an in-house distributed solution called Node
Scala~\cite{Maatouki:2015aa}. The reason for this is that while
undoubtedly optimised for performance, Node.js applications are by
design restricted to a single thread. Distributed systems featuring
multiple instances of the same application accessed through a common
interface such as the same HTTP server, can be constructed using the
standard Node.js module Cluster (on a single host) or third-party
solutions built on top of it
such as StrongLoop Process Manager (which can support multiple hosts),
however neither of these solutions allow for parallel processing.
Conversely, Node Scala
does allow parallel execution of tasks with intelligent distribution
of chunks among back-end servers (\textit{i.e.} workers) and supports
the spawning and monitoring of back-end servers on both the same and
multiple hosts. Please see~\cite{Maatouki:2015aa} for more information
about Node Scala and its performance.

In its current configuration our Node Scala instance runs on a single,
dedicated server. It uses a single front-end server offering the MIPAS
REST API over HTTPS and two back-end servers which handle
communication with the database server as well as data
processing. Following the rule of least privilege, back-end servers
are only allowed to read from the database.

\subsection{The Client}
\label{sec:sysArchClient}

Our new client application, KAGLVis, is a data browser which displays
selected observables as 3D points at correct coordinates on a virtual
globe. At present it can display orbital paths of Envisat as well as
cloud altitude measured by MIPAS, in the latter case allowing the user
to specify criteria defining clouds. The data is fetched from the
server in the background, depending on the user's preferences either
set by set or simultaneously for the whole selected range. The view,
which can be either a sphere, a plane or that of poles and which
allows for selection of a number of different Earth images as
background, can be freely rotated and zoomed. The colour map in the legend is drawn
dynamically on a HTML5 canvas element and synchronised with the
contents of the 3D view.

It is worth emphasising at this point that KAGLVis is \emph{not} meant
as a replacement for the plethora of standard data-processing tools
used in geosciences. Instead, KAGLVis aims
to provide basic visualisation and analysis capabilities as easily as
possible.

The internal logic of KAGLVis has been implemented using AngularJS,
with each component of the view assigned its own controller. All
communication between components occurs explicitly through messages
sent via a dedicated internal service; no communication through
\textit{e.g.} global variables is allowed. Another service provides
access to configuration of the application. Finally, the third service
provides an interface to the data source --- which by default is our
REST server but can if need be, for testing for instance, switched to
a local JSON file.

The heart of KAGLVis, the 3D display, is based on WebGL Globe --- a
lightweight JavaScript virtual globe created by Google Data Arts
Laboratory which can display longitude-latitude data as
spikes. As the name suggests, Globe uses WebGL to
leverage local GPU power for 3D rendering. Our version of Globe has
been customised to support other views than the original sphere as
well as selection of the background texture, reduce memory consumption
at a cost of disabling certain visual effects, and most importantly to
allow caching of previously displayed data sets should the user want
to return to them at some point.

\begin{figure}
  \centering
  \includegraphics[width=\columnwidth]{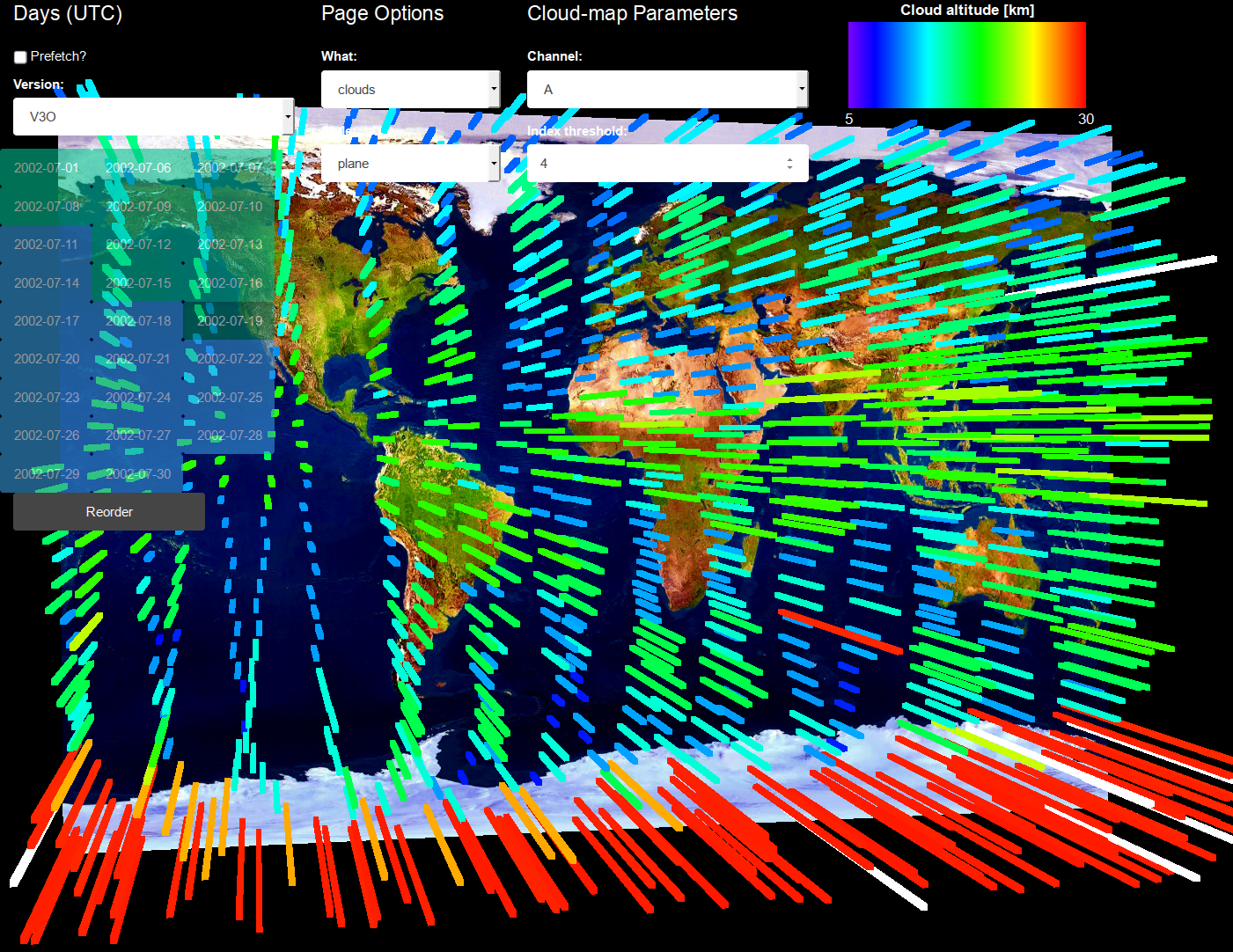}
  \caption{A screenshot of KAGLVis showing cloud altitude measured by
    MIPAS against flat Earth view.}
  \label{fig:screenshotKAGLVis}
\end{figure}

Testing has shown KAGLVis is indeed highly portable --- it has been
confirmed to run without errors on many Web browsers and operating
systems, including on mobile devices.

We use a minimal Express-based Web server to serve KAGLVis to
users. The server is hosted on the same system as the REST server because,
as a single-page application, KAGLVis adds only minimal load to the server hosting it.

\section{Evaluation}
\label{sec:Evaluation}

We have measured performance and resource use of KAGLVis under
realistic operating conditions. There are three reasons for focusing
on the client: it is the component through which our system is
experienced by most users, Web browsers are not as strongly optimised
for performance as MongoDB and Node.js, and some benchmarks of
back-end components of our system have already been published (see
\textit{e.g.}~\cite{Ameri:2014aa, Maatouki:2015aa}).

Our tests were executed on a PC with
an Intel Core i5-4300U CPU with HD 4400 graphics, 12~GB of DDR3/1600 RAM,
and a LCD screen at its native
resolution of 1920x1200 pixels at 60~Hz.
The system ran an up-to-date 64-bit installation of Gentoo
Linux, including X.org
server 1.16.4 and the video driver xf86-video-intel
2.99.917. All tests were run in the Chromium 44.0.2403.89 Web browser
using one of its built-in developer tools, Timeline~\cite{web:ChromeTimeline}.
The client connected to our production REST server (the technical
details of which can be found in \cite{Maatouki:2015aa}),
using IPv4 over a Gigabit
Ethernet connection.

We benchmarked the visualisation of cloud altitude measured by MIPAS.
This application relies on user input and thus requires some
client-side processing of data.  For each test and iteration we
launched Chromium fresh and recorded Timeline events from KAGLVis for
up to 20 days' worth of data, a number high above typical usage patterns ---
already at 5 days parts of the map become too crowded to be read
comfortably.

\begin{figure}
  \centering
  \includegraphics[width=\columnwidth]{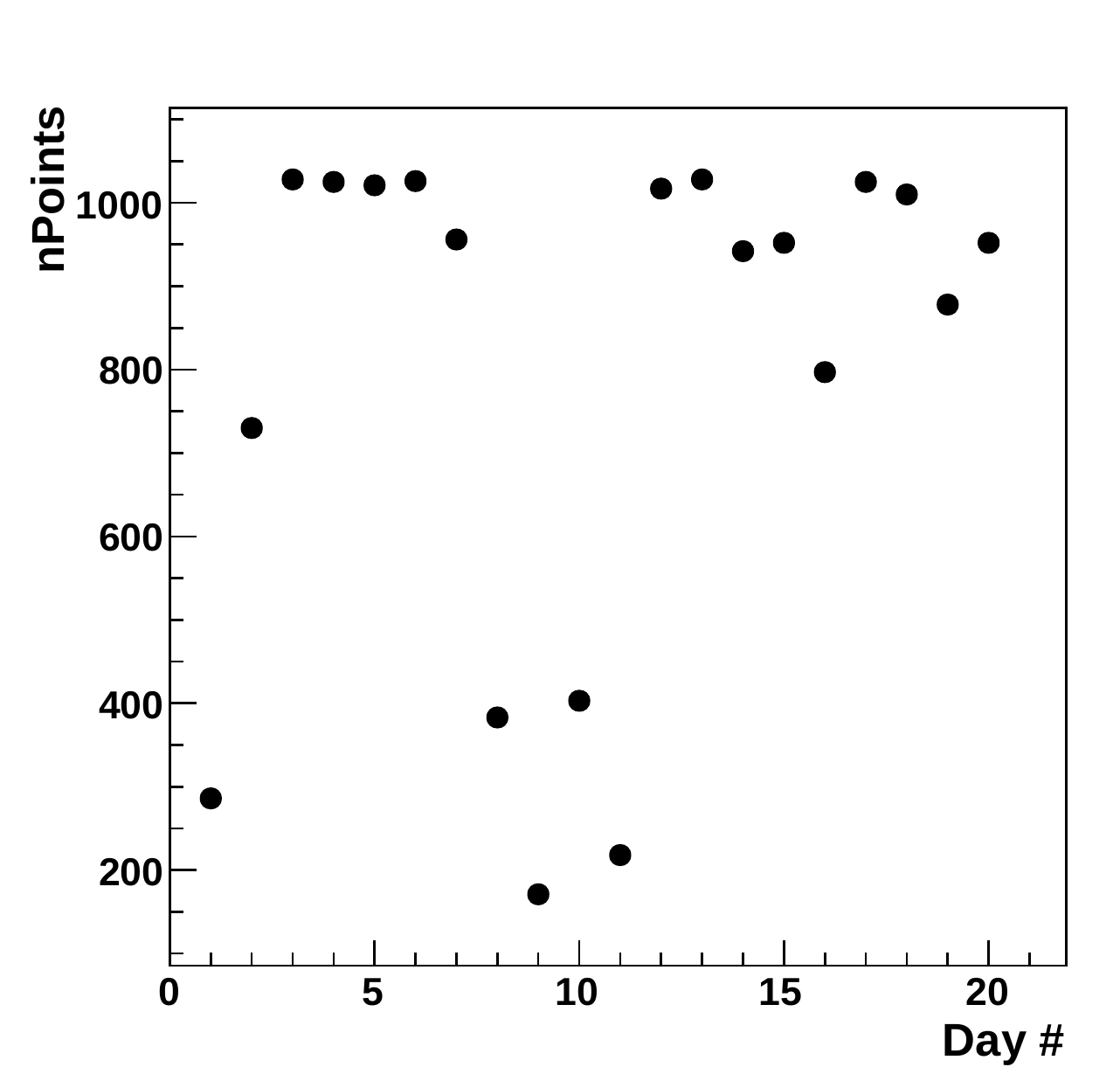}
  \caption{Number of data points for each day of Envisat MIPAS data
    used in subsequent tests. Horizontal axis: index of the day in the
    complete test sample. Vertical axis: number of data points the day
    contains.}
  \label{fig:nPoints}
\end{figure}

The number of usable MIPAS data points per day varies. Figure~\ref{fig:nPoints}
shows the number of points for each day displayed during testing. In
order to account for differences in data-set size we, where
appropriate, normalised results of our measurements to the number of
points in each day.

\subsection{Time to Display}
\label{sec:timeToDisplay}

To measure how responsive KAGLVis it is to input, we first measured
how long takes to display a new data set on the screen ---
\textit{i.e.} convert JSON input to a 3D mesh, add that mesh to the
WebGL scene and update the view to reflect the changes --- after it
has been requested by the user. Subsequent days are combined with
previously displayed content instead of replacing it so that we can
watch for possible scalability problems. We explicitly excluded
data-transfer time from the benchmark because network transfers in
KAGLVis occur asynchronously in the background, not affecting
interface responsiveness.

\begin{figure}
  \centering
  \includegraphics[width=\columnwidth]{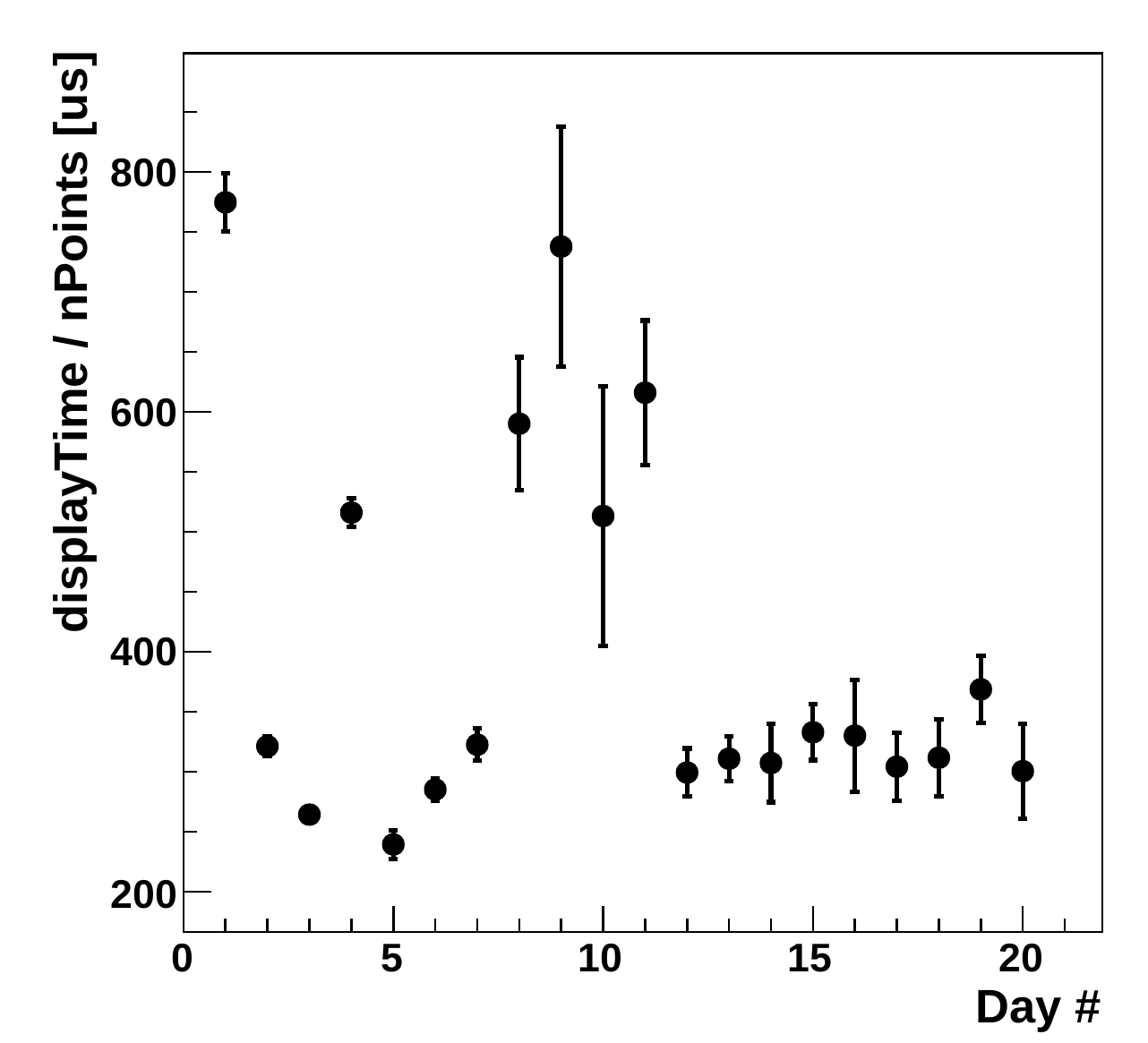}
  \caption{Time required to add a day's worth data to the
    view. Horizontal axis: index of the day in the complete test
    sample. Vertical axis: time required to display the day divided by
    the number of data points it contains.}
  \label{fig:timeToDisplay}
\end{figure}

As Figure~\ref{fig:timeToDisplay} shows, for most data sets it
consistently takes around 300~$\mathrm{\mu s}$ per point, or 0.3~s
whole, to display.  The exception, day 4, seems to be caused by the
WebGL implementation in Chromium --- Timeline shows that while the
time required to add this day to the scene is consistent with that for
other data sets of similar size, the subsequent animation-frame update
not only takes considerably longer than for all other data sets but is
in fact split into three separate steps.

For days low-point days 8--11 the behaviour is different --- around
600~$\mathrm{\mu s}$ per data point (amounting to 1--2~s per whole
data set), with considerable fluctuations between iterations. Given
the detailed structure of Timeline events for these and other days
appears to be very similar, it is believed that for data sets so small
the measured time becomes dominated by the processing overhead.

Finally, for day 1 the time of almost 780~$\mathrm{\mu s}$ per point
is even slightly higher, while fluctuating less, than for other small
data sets As the Timeline structure of the animation-frame update step
here is very different from that for subsequent data sets, we
concluded what we see here is both processing overhead and effects of
WebGL initialisation.

\subsection{Memory Consumption}
\label{sec:memoryConsumption}

\begin{figure}
  \centering
  \includegraphics[width=\columnwidth]{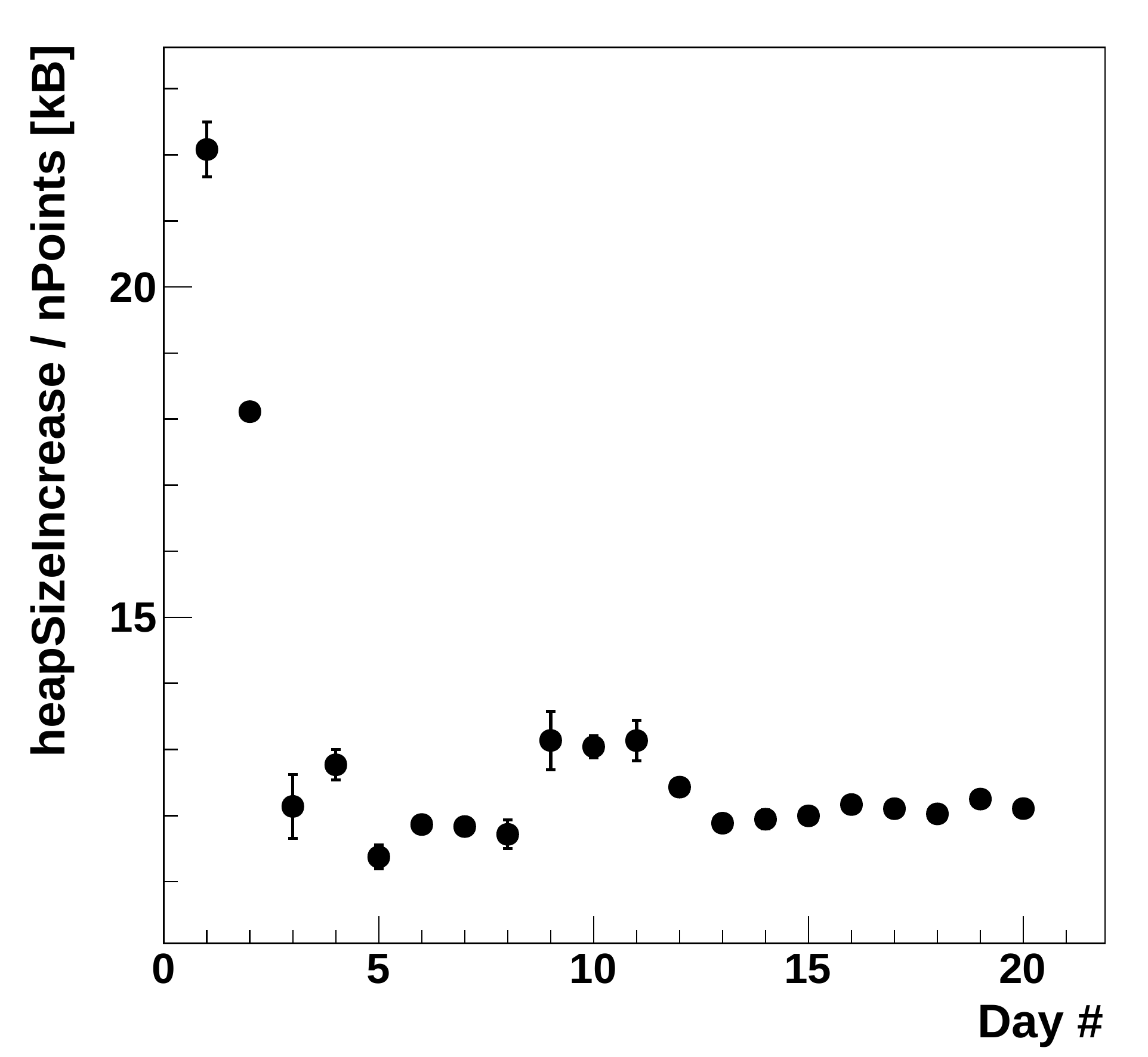}
  \caption{Evolution of memory consumption as more and more data is
    added to the view. Horizontal axis: index of the day in the
    complete test sample. Vertical axis: difference in JavaScript heap
    size after another day's worth of data has been displayed and
    before it has been loaded into the view, divided by the number of
    data points in each day.}
  \label{fig:heapUse}
\end{figure}

Secondly, we measured the growth of the
JavaScript heap of KAGLVis as we load more and more data sets into it.
The results can be found in Figure~\ref{fig:heapUse}.

A consistent increase of around 12~kB per data point can be observed
from day 3 onward --- suggesting the test system could cache hundreds
of days' worth of data before running out of physical
RAM. Above-average value of 22~kB per point measured for day 1 is
consistent with our earlier hypothesis of the loading of the first
data set triggering initialisation of internal data structures. It is
presently not fully understood why an above-average value can also be
observed for day 2.

\subsection{Frame Rate}
\label{sec:frameRate}

Finally, we used the FPS counter built into Chromium to measure the
frame rate of KAGLVis window as a function of the number of
simultaneously shown data sets. As integrated graphics chipsets are
generally not performance-oriented, we repeated this test on an
otherwise similar PC with an AMD Radeon HD 4770 PCIe graphics
card. Apart from the graphics driver (Mesa-10.3.7 with Gallium3D
driver ``r600'' + X.org driver xf86-video-ati 7.5.0) the software on
both systems was identical.

What makes this comparison particularly interesting is that we are
comparing a relatively recent Intel graphics chipset with a device
which while originally marketed as mid-range is now quite old --- the
two were released in late 2013 and early 2009, respectively.

\begin{figure}
  \centering
  \includegraphics[width=\columnwidth]{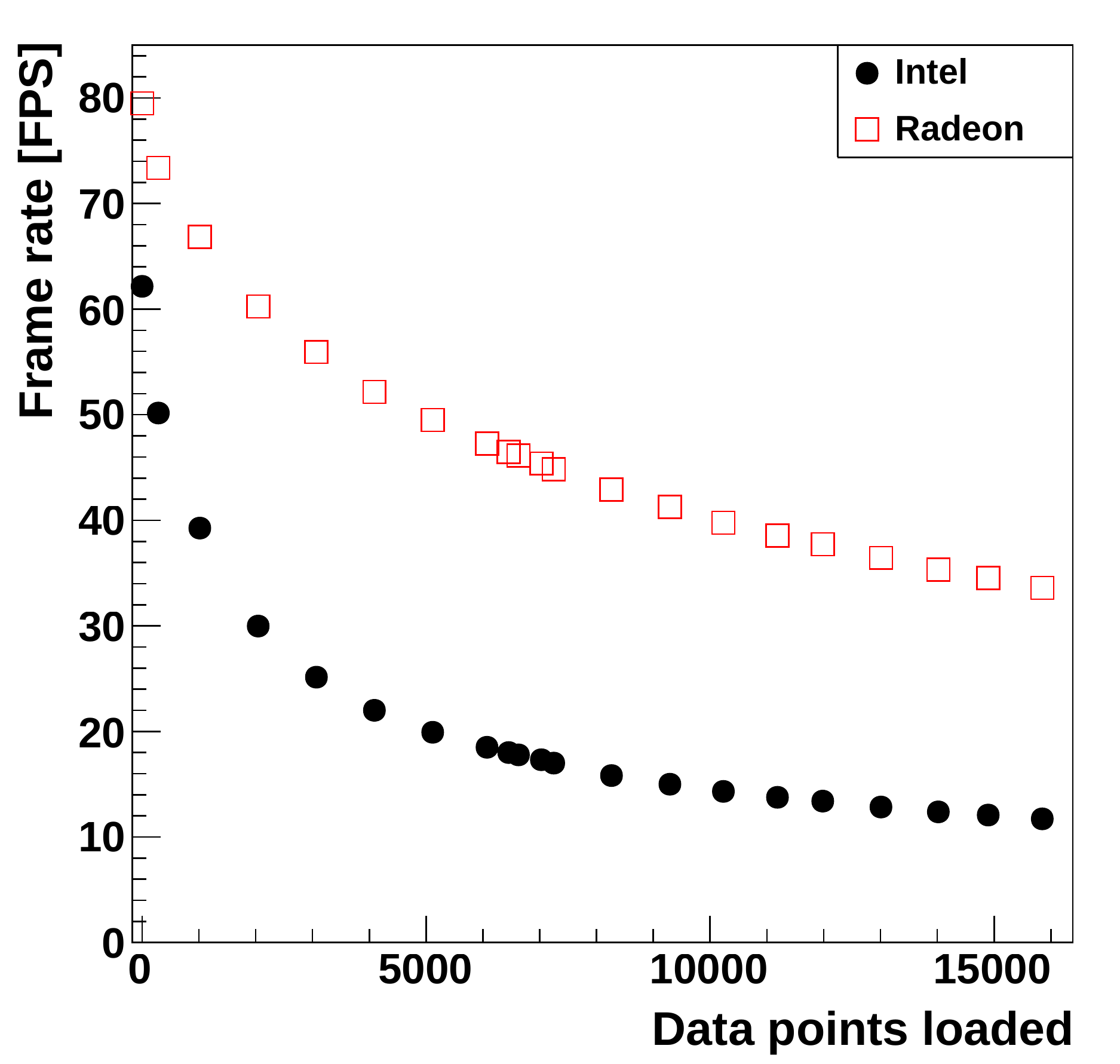}
  \caption{Frame rate as a function of the number of data points
    simultaneously displayed on screen. Black: results from our
    standard test system. Red: results from a machine with AMD Radeon
    graphics.}
  \label{fig:fps}
\end{figure}

Measurements were repeated five times for
each set-up and their results can be found in Figure~\ref{fig:fps}.

The frame rate drops dramatically as more data is added to the
view. With all points visible it is just under 12~FPS, resulting in
noticeable jerkiness of animation during rotation or zooming. That
said, frame rates in the typical use range of under 5,000 points
appear to be sufficient.

Despite its age the Radeon card outperforms the Intel chipset: around
34~FPS for all points and above 50~FPS in the typical use range.

A number of peculiarities can be observed in the results. To begin
with, the frame-rate drop does not scale linearly with the number of
points. Instead, it becomes smaller and smaller as more data is added
to the view. It is furthermore interesting that such low frame rates
are seen for a fairly uncomplicated and essentially static scene,
especially given the very same system consistently outputs more than
30~FPS running a highly dynamic and complex WebGL water
simulation. All in all it would seem that the observed behaviour is
driven primarily by the properties of the WebGL engine provided by
Chromium, although given the behaviour of the aforementioned water
test optimisation of WebGL use in Globe might improve its 3D
performance as well.

\section{Related Work}
\label{sec:relatedWork}

Given both the number of different components a system capable of
comprehensive handling of geospatial data must contain and how
widespread the handling of such data has become in modern science, it
is not surprising the amount of work put into this topic has been
considerable. Here are some examples.

On the database side one should definitely mention
PostGIS~\cite{DBLP:reference/gis/Strobl08a}, a spatial database
extension for the PostgreSQL relational database system which is
supported by a wide range of GIS applications. However, as a RDBMS
it is not in direct competition with NoSQL-based systems like ours.

Even if only Node.js-based solutions are considered, the number of
alternatives for server-side code is considerable. Then again, each of
them has certain shortcomings --- which ultimately led us to develop
Node Scala. Please see~\cite{Maatouki:2015aa} for details on this
matter.

Likewise there are many solutions for visualisation of geospatial data
and each of them has disadvantages from our perspective. For example,
the popular and open NASA World
Wind~\cite{Hogan:2011:NWW:1999320.1999322} is presently only suitable
for standalone applications and would introduce another programming
language (Java) into the stack, while at the same time a lot of its
features and detail are simply unnecessary while dealing with
high-altitude data from satellites. Another alternative, Google Earth
Engine~\cite{2013EGUGA..1511997G}, is not unlike our own system in
that it is an all-in-one solution handling both storage,
data-management and analysis through a highly distributed Web
application --- but does not presently give most of its users the
possibility to develop own processing algorithms, is a closed platform
tied to Google computing infrastructure, and most importantly can only
be used to process data made available by Google.

Another example is the highly popular UltraScale Visualisation Climate
Data Analysis Tools (UV-CDAT) framework~\cite{10.1109/MC.2013.119}.
UV-CDAT and our system complement rather than compete with each other:
the latter attempts first and foremost to provide high-performance
access to data, the latter focuses on data analysis.

Finally, the German Satellite Data Archive (D-SDA) of the Earth
Observation Center at the German Aerospace Center (DLR)~\cite{Reck:2011aa}
at a glance seems to
serve the same purpose as our system yet is backed by considerably
more resources. There are, however, differences: it is tied to DLR
infrastructure, keeps both data and metadata
as files in storage rather than in a database, and most importantly
follows considerably different philosophy --- its primary goal is to
achieve long-term data preservation over more than 20 years with
nearly exponentially growing data capacity, whereas our archive has
been designed primarily to provide high-performance access to a fixed
chunk of data. One could therefore imagine the two systems as
complementary --- ours for rapid application-specific processing,
D-SDA for long-term storage.

\section{Conclusions and Future Work}
\label{sec:Conclusions}

We have demonstrated a distributed, high-performance and scalable
system based on the so-called MEAN stack, which is used to store,
process and visualise data from the ESA Envisat Earth-observing
satellite. Benchmarks of the system as seen from the end-user's
perspective demonstrating good performance even beyond the typical use
range and on a system without a discrete graphics device.

In the near future we will continue importing further MIPAS data
into MongoDB
as well as prepare for the migration of the production database to a
sharded cluster. The REST server shall be extended accordingly, with
new use cases and more back-end servers. Finally, we would like to
improve KAGLVis user experience on mobile devices.

\section*{Acknowledgements}

This work is funded by the project ``Large-Scale Data Management and
Analysis''~\cite{Jung:2014oba}, funded by the German Helmholtz
Association.

\bibliographystyle{IEEEtran}
\bibliography{references}

\begin{thebibliography}{10}
\providecommand{\url}[1]{#1}
\csname url@samestyle\endcsname
\providecommand{\newblock}{\relax}
\providecommand{\bibinfo}[2]{#2}
\providecommand{\BIBentrySTDinterwordspacing}{\spaceskip=0pt\relax}
\providecommand{\BIBentryALTinterwordstretchfactor}{4}
\providecommand{\BIBentryALTinterwordspacing}{\spaceskip=\fontdimen2\font plus
\BIBentryALTinterwordstretchfactor\fontdimen3\font minus
  \fontdimen4\font\relax}
\providecommand{\BIBforeignlanguage}[2]{{%
\expandafter\ifx\csname l@#1\endcsname\relax
\typeout{** WARNING: IEEEtran.bst: No hyphenation pattern has been}%
\typeout{** loaded for the language `#1'. Using the pattern for}%
\typeout{** the default language instead.}%
\else
\language=\csname l@#1\endcsname
\fi
#2}}
\providecommand{\BIBdecl}{\relax}
\BIBdecl

\bibitem{Garcia:2011aa}
A.~Garcia, S.~Bourov, A.~Hammad, J.~van Wezel, B.~Neumair, A.~Streit,
  V.~Hartmann, T.~Jejkal, P.~Neuberger, and R.~Stotzka, ``{The Large Scale Data
  Facility: Data Intensive Computing for Scientific Experiments},'' in
  \emph{Parallel and Distributed Processing Workshops and PhD Forum (IPDPSW),
  2011 IEEE International Symposium on}, May 2011, pp. 1467--1474.

\bibitem{web:MEANstack}
\BIBentryALTinterwordspacing
V.~Karpov. {The MEAN Stack: MongoDB, ExpressJS, AngularJS and Node.js}. (last
  visited: 2015-11-06). [Online]. Available:
  \url{http://blog.mongodb.org/post/49262866911}
\BIBentrySTDinterwordspacing

\bibitem{Ameri:2014aa}
P.~Ameri, U.~Grabowski, J.~Meyer, and A.~Streit, ``{On the Application and
  Performance of MongoDB for Climate Satellite Data},'' in \emph{{Proceedings
  of the 13th IEEE International Conference on Trust, Security and Privacy in
  Computing and Communications}}, 2014, pp. 652--659.

\bibitem{Maatouki:2015aa}
A.~Maatouki, M.~Szuba, J.~Meyer, and A.~Streit, ``{A horizontally-scalable
  multiprocessing platform based on Node.js},'' in \emph{Proceedings of the
  13th IEEE International Symposium on Parallel and Distributed Processing with
  Applications (to be published)}, 2015, arXiv:1507.02798 [cs.DC].

\bibitem{web:ChromeTimeline}
\BIBentryALTinterwordspacing
{Performance profiling with the Timeline}. (last visited: 2015-11-06).
  [Online]. Available:
  \url{https://developer.chrome.com/devtools/docs/timeline}
\BIBentrySTDinterwordspacing

\bibitem{DBLP:reference/gis/Strobl08a}
\BIBentryALTinterwordspacing
C.~Strobl, ``{PostGIS},'' in \emph{Encyclopedia of {GIS.}}, S.~Shekhar and
  H.~Xiong, Eds.\hskip 1em plus 0.5em minus 0.4em\relax Springer, 2008, pp.
  891--898. [Online]. Available:
  \url{http://dx.doi.org/10.1007/978-0-387-35973-1\_1012}
\BIBentrySTDinterwordspacing

\bibitem{Hogan:2011:NWW:1999320.1999322}
\BIBentryALTinterwordspacing
P.~Hogan, ``{NASA World Wind: Infrastructure for Spatial Data},'' in
  \emph{Proceedings of the 2nd International Conference on Computing for
  Geospatial Research \& Applications}, ser. COM.Geo '11.\hskip 1em plus 0.5em
  minus 0.4em\relax New York, NY, USA: ACM, 2011, pp. 2:1--2:1. [Online].
  Available: \url{http://doi.acm.org/10.1145/1999320.1999322}
\BIBentrySTDinterwordspacing

\bibitem{2013EGUGA..1511997G}
N.~{Gorelick}, ``{Google Earth Engine},'' in \emph{EGU General Assembly
  Conference Abstracts}, ser. EGU General Assembly Conference Abstracts,
  vol.~15, Apr. 2013, p. 11997.

\bibitem{10.1109/MC.2013.119}
D.~N. Williams, T.~Bremer, C.~Doutriaux, J.~Patchett, S.~Williams, G.~Shipman,
  R.~Miller, D.~R. Pugmire, B.~Smith, C.~Steed, E.~W. Bethel, H.~Childs,
  H.~Krishnan, P.~Prabhat, M.~Wehner, C.~T. Silva, E.~Santos, D.~Koop,
  T.~Ellqvist, J.~Poco, B.~Geveci, A.~Chaudhary, A.~Bauer, A.~Pletzer,
  D.~Kindig, G.~L. Potter, and T.~P. Maxwell, ``{Ultrascale Visualization of
  Climate Data},'' \emph{Computer}, vol.~46, no.~9, pp. 68--76, 2013.

\bibitem{Reck:2011aa}
C.~Reck, E.~Mikusch, S.~Kiemle, K.~Molch, and W.~Wildegger, ``{Behind the
  Scenes at the DLR National Satellite Data Archive, a Brief History and
  Outlook of Long Term Data Preservation},'' in \emph{{Proceedings, PV 2011:
  Ensuring Long-Term Data Preservation, and Adding Value to Scientific and
  Technical Data}}, Toulouse, France, Nov. 2011, http://elib.dlr.de/74103/.

\bibitem{Jung:2014oba}
C.~Jung, M.~Gasthuber, A.~Giesler, M.~Hardt, J.~Meyer, F.~Rigoll, K.~Schwarz,
  R.~Stotzka, and A.~Streit, ``{Optimization of data life cycles},'' \emph{J.
  Phys. Conf. Ser.}, vol. 513, p. 032047, 2014.

\end{thebibliography}

\end{document}